\documentclass{lmcs}

\usepackage{amsmath,amssymb}
\usepackage{listings}

\newcommand{\Om}{\Omega}
\newcommand{\PSh}{\mathbf{PSh}}
\newcommand{\Set}{\mathbf{Set}}
\newcommand{\dom}[2]{\mathrm{do}(#1\!:=\!#2)}
\newcommand{\cl}{\bigcirc}
\newcommand{\nn}{\neg\neg}

\makeatletter
\def\endfront@text{%
\begin{figure}%
\hsize\textwidth%
\fontsize{6}{7\p@}\normalfont\upshape%
\noindent\hfill
\hbox{\fontencoding{T1}%
  \vbox{\fontsize{6}{8 pt}\baselineskip=6 pt\vss
    \hbox{\headertextsf{\hspace{2 pt}\copyright\quad \shortauthors}\hfil}
    \hbox{\headertextsf{\crC\ \href{http://creativecommons.org/about/licenses}{Creative Commons}}\hfil}}}%
\par\kern\z@
\end{figure}%
}
\makeatother

\begin{document}

\title[A cubical formalisation of topos causal models]{A cubical formalisation of topos causal models: \\ intervention, forcing, and a contextuality obstruction}

\author[K.~Sargsyan]{Karen Sargsyan}
\address{Institute of Chemistry, Academia Sinica, Taipei, Taiwan}
\email{karen.sarkisyan@gmail.com}

% First-page classification footnote suppressed (keywords / MSC / ACM CCS not rendered).
% Re-enable for journal submission if required. Note: the MSC list is MSC2020
% (68V20, the proof-assistant/formalization class, was introduced in MSC2020),
% and lmcs.cls hard-codes the label "2010", so also override \amsclassname if re-enabling.
% \keywords{cubical type theory, topos causal models, subobject classifier, sheaf, Kripke-Joyal semantics, Lawvere-Tierney topology, do-calculus, formal verification}
% \amsclass{68V20, 18B25, 03G30, 68Q55, 03B70}
% \ACMCCS{Theory of computation~Type theory; Theory of computation~Categorical semantics; Mathematics of computing~Causal networks}

\begin{abstract}
Topos causal models recast causal inference inside a topos: a causal world is a presheaf, an intervention is a sub-model named by a characteristic map into the subobject classifier $\Om$, and reasoning is Kripke-Joyal forcing in an intuitionistic internal language. We give the first axiom-free machine-checked account of this 1-topos core, in Cubical Agda over a previously verified probability monad and do-calculus; the framework is otherwise developed on paper, with central claims stated rather than proved. Three of our results go beyond faithful transcription. We exhibit a contextuality obstruction the programme does not treat: pairwise-consistent local causal data with no global model, detected by a degree-one holonomy class. We delimit the claim that interventions are modelled by the subobject classifier: an intervention and an observation name the same subobject, so $\Om$ fixes the target of a do-operation but not the operation itself, which is surgery on the kernels --- where, on a confounder, the interventional and observational laws differ. And we settle the modal unit --- inflationarity is derivable from $j\top = \top$ and naturality, not a fourth axiom. We also machine-check the classifier of sieves with its classification theorem, the pullback collating local mechanisms, and the Kripke-Joyal forcing clauses. The development assumes no axioms and typechecks under Agda's \texttt{--safe} flag, with the ordered field discharged at $\mathbb{Q}$; type-level sheafification and a directed do-calculus are future work.
\end{abstract}

\maketitle

\section{Introduction}\label{sec:intro}

A structural causal model is a system of equations: each variable is a function of its causes and an independent noise term, and an intervention replaces one equation by a constant assignment.
Mahadevan's \emph{topos causal models} (TCMs)~\cite{mahadevan-2025a,mahadevan-2025c} recast this picture categorically.
A causal world is a presheaf over a base category of contexts; a mechanism is a morphism of presheaves; and an intervention is no longer an edit to a syntax tree but a \emph{sub-model}, a monic arrow $M_x \hookrightarrow M$ into its parent model, named by its characteristic map into the subobject classifier $\Om$, the object of truth values of the topos.
Because a presheaf topos has an intuitionistic (Heyting, not Boolean) internal logic, causal statements live in that internal language, interpreted by Kripke-Joyal forcing.
A companion line of Mahadevan's work~\cite{mahadevan-2025b} adds an \emph{intuitionistic $j$-do-calculus}, in which a Lawvere-Tierney topology $j : \Om \to \Om$ acts as a modality picking out the causal statements that are stable under localisation.

The appeal of this reformulation is that several pieces of causal machinery become standard topos-theoretic universal properties: intervention is a subobject, named by a characteristic map into $\Om$; the assembly of local mechanisms into a global one is gluing; and counterfactual reasoning is forcing in the internal logic.
Two further phenomena live on the same machinery: the $j$-do-calculus modality that isolates the causal facts invariant across regimes, and --- new to this account --- a cohomological obstruction, the failure of pairwise-consistent local models to assemble into a global one.
The framework is, however, developed on paper, and some of its central claims are stated rather than proved.

This paper supplies that verification: a machine-checked account of the 1-topos core in Cubical Agda~\cite{cchm-2018,cubical-agda}.
It is a direct sequel to a companion development~\cite{sargsyan-cubical-dsep}, which verifies a probability monad as a higher inductive type together with Pearl's do-calculus rules and the soundness of d-separation (the graphical criterion for which conditional independences a DAG entails); that development supplies the mechanisms (probability kernels) that the present one organises into a topos.
Throughout we work in the presheaf topos $\PSh\,\mathcal{C} = \Set^{\mathcal{C}^{\mathrm{op}}}$ over an abstract base category $\mathcal{C}$ of contexts, exactly the setting in which Mahadevan's general definition of $\Om$ lives.

\paragraph{Contributions}
Our account is faithful to the source where the source is precise, and in three places it does more than transcribe: it delimits the account of intervention --- the classifier names the target, not the operation --- settles the status of the modal unit (it is derivable, not a missing axiom), and adds an obstruction the programme leaves out.
\begin{itemize}
\item \emph{The intervention's target and its classifier} (Section~\ref{sec:classifier}). We build $\Om$ as the presheaf of sieves and realise the value-fixing subobject ``$X = x_0$'' as the characteristic map $\chi : X \Rightarrow \Om$. We prove the classification theorem $\chi_c(b) = \top \iff b = x_0\,c$ in both directions, and connect $\chi$ to the operational, kernel-level intervention of the companion paper. We also prove the general universal property --- every restriction-closed sub-presheaf is classified by a \emph{unique} map into $\Om$, the standard characterisation of the subobject classifier --- and the value-fixing map is one instance. A small design point falls out: a topos-internal intervention must fix $X$ to a \emph{natural} global element, not merely a context-indexed family.
\item \emph{What the classifier does not classify} (Section~\ref{sec:dosee}). The programme's account of intervention is that a sub-model is a monic arrow into its parent model, classified by $\Om$. We delimit it: at the level of $\Om$ the event ``$X = x_0$'' is the same whether it is intervened or observed, so no map into $\Om$ separates surgery from conditioning. We machine-check that the two do differ --- on a confounder the interventional and observational laws of the outcome are distinct (module \texttt{DoSeeDistinct}), with a concrete interior witness so the statement is not vacuous --- and that the surgery pins $X$ to the value $\chi$ classifies. The classifier names the target of an intervention; the kernels beneath it carry the operation. This is a limit on what the subobject classifier contributes, not a defect in the construction.
\item \emph{Collation by pullback} (Section~\ref{sec:gluing}). Mahadevan motivates sheaves with the claim that local mechanisms ``collate'' into a unique global one, but proves no existence theorem for collation: the companion's local-to-global result assumes a sheaf of causal kernels and gives only uniqueness. We verify the limit form: two mechanisms agreeing on a shared target have a pullback, computed contextwise, with its universal property. Collation over a cover is instead the sheaf condition, which a presheaf topos does not satisfy in general and which we do not supply: we neither formalise sheafification nor assume it compatible with our distribution carrier (Section~\ref{sec:scope}). And it is that collation over a cover --- not the pullback, which always exists --- that three contexts can obstruct entirely (Section~\ref{sec:obstruction}).
\item \emph{The internal language} (Section~\ref{sec:forcing}). We define Kripke-Joyal forcing and machine-check the clause for each connective and quantifier: $\top, \bot, \wedge, \vee, \Rightarrow, \forall, \exists$, together with the locality (monotonicity) property that defines a Kripke semantics.
\item \emph{An instantiated do-calculus, the modal unit settled} (Section~\ref{sec:modal}). We machine-check that inflationarity $S \le j\,S$ --- the unit of the modality --- is \emph{derivable} from $j\top = \top$ and naturality, so a Lawvere-Tierney topology on $\Om$ needs no fourth axiom (module \texttt{InflationarityDerivable}). The three equations \emph{alone} do not force it: on a bare three-element chain with its meet and induced order, where naturality has no meaning, the equations hold yet a truth value is collapsed (module \texttt{InflationarityIndependence}); the derivation therefore rests on naturality. We exhibit the double-negation topology $\nn$ as a concrete non-degenerate instance (the source never specialises $j$ on $\Om$ to a concrete operator; its instances are Grothendieck covers of regimes), present the modality $\cl = j$ as a closure operator on truth values --- its modal objects a sub-poset of $\Om$, not a subuniverse of types --- and prove that the value the intervention forces, and Pearl's Rules~1, 2 and~3, are $j$-closed for every topology, and hence stable under any sheaf localisation.
\item \emph{An obstruction} (Section~\ref{sec:obstruction}). The pullback always exists, but collation over a cover can fail: we machine-check that pairwise-consistent local causal data may admit no global model --- strong contextuality, formalised combinatorially --- a phenomenon outside the programme's scope (Section~\ref{sec:related} positions it against prior contextuality work).
\end{itemize}
The whole development typechecks under Agda's \texttt{--safe} flag --- no postulates, holes, or unsafe features anywhere in it --- with the underlying ordered field discharged concretely at $\mathbb{Q}$. It is openly available at \url{https://github.com/karsar/cubical-topos-causal}.

\paragraph{Significance}
Mechanising a paper framework is worthwhile in two ways here.
One is assurance: the framework's central claims are now certified rather than asserted, which matters for a programme aimed at automated causal reasoning~\cite{mahadevan-catagi-2026}, where a verified core is what machine-generated models are checked against.
The other is that carrying out the proofs brought out two points that matter in practice.
First, the modal layer's guarantees rest on $j$ being a closure operator; on the classifier this is automatic, since naturality and $j\top = \top$ supply the unit.
What the formalisation establishes is that this is a property of the classifier, not of the three equations: a formalisation that models topologies on a bare Heyting algebra would silently lose it (Section~\ref{sec:modal}).
Second, the gluing on which modular, map-reduce assembly of local mechanisms rests can be obstructed --- over three contexts, locally consistent pieces can share no global model at all (Section~\ref{sec:obstruction}).
The practical implication is that modular causal discovery needs a global-consistency check, not merely pairwise agreement; and because the obstruction is cohomological, that check is in principle computable.
We verify it on a minimal three-context example rather than on data; what this establishes is that the failure mode is real and exactly located, not that it is common.

\paragraph{Scope}
This is the \emph{1-topos} (presheaf) account: truth values are sieves, the modality acts on truth values, and the internal logic is the standard Kripke-Joyal one.
We do not formalise type-level sheafification (the associated-sheaf reflector on the whole topos), which is a larger construction whose effect on our distribution carrier we leave open, nor the \emph{directed} lift of the do-operator itself, which needs directed univalence and so lies outside current tools; its structural core, however, we do machine-check in \texttt{rzk} (Section~\ref{sec:scope}).
Section~\ref{sec:scope} states the boundary precisely.

\section{The presheaf topos and its subobject classifier}\label{sec:background}

We fix a base category $\mathcal{C}$ of \emph{contexts} (Mahadevan's ``regimes''~\cite{mahadevan-2025b}): objects are contexts, morphisms are admissible context maps.
A \emph{causal world} is a presheaf $X : \mathcal{C}^{\mathrm{op}} \to \Set$: a family of value-sets $X(c)$ with a restriction map $X(f) : X(c) \to X(d)$ for each $f : d \to c$, functorial in $f$.
We write composition in diagrammatic order, $f \star g$ meaning ``$f$ then $g$'', so that restriction composes left to right.
A \emph{mechanism} is a natural transformation $X \Rightarrow Y$.

\paragraph{The classifier of sieves}
A \emph{sieve} on a context $c$ is a downward-closed family of morphisms into $c$: a predicate $S$ on arrows $f : d \to c$ such that $f \in S$ implies $k \star f \in S$ for every $k$.
Sieves on $c$ form a set, and pulling a sieve back along $h : c' \to c$ (keeping the arrows $f$ with $f \star h \in S$) makes the assignment $c \mapsto \{\text{sieves on } c\}$ a presheaf $\Om$.
Over the two-object poset $0 \to 1$, for instance, there are two sieves on $0$ but \emph{three} on $1$: the empty sieve, the single arrow $\{0 \to 1\}$, and the maximal sieve. That middle one --- true only after restriction to $0$ --- already shows the non-Boolean, intuitionistic character of $\Om$.
This is the \emph{subobject classifier}: a sub-presheaf $A \hookrightarrow B$ is named by a unique map $\chi : B \Rightarrow \Om$, with $b \in A$ exactly when $\chi(b)$ is the maximal sieve $\top$ (all arrows).
This is precisely Mahadevan's definition (his sieves are subobjects of the representable $\mathcal{C}(-,x)$, and his $\Om(X)$ is their collection~\cite[Def.~8,11]{mahadevan-2025c}); we restate it as an Agda record and verify the presheaf laws.
Membership of a sieve is a proposition, which keeps $\Om(c)$ a set --- needed for $\Om$ to be a presheaf of sets --- and lets us reason about sieve equality by their membership alone.

\paragraph{Internal mechanisms}
The mechanisms we intervene on come from the companion paper~\cite{sargsyan-cubical-dsep}.
There, a two-variable structural model on sets $X,Y$ is a prior $p_X$ on $X$ together with a kernel $k_Y : X \to \mathtt{FDist}\,Y$ into the probability monad, and the intervention $\dom{X}{x_0}$ replaces $p_X$ by the point mass at $x_0$.
Internally, a model on presheaves $X, Y$ is such a model contextwise --- one at each context $c$ --- and the internal intervention is the contextwise point-mass substitution.
The marginal of the outcome $Y$ is a context-indexed element of the internal distribution object.
We take all of this as given; the present paper studies how interventions sit inside the topos and, in Section~\ref{sec:dosee}, what at the kernel level separates an intervention from an observation.

\section{The intervention's target and its classifier}\label{sec:classifier}

Fix a value presheaf $X$ and a value $x_0$ to which we wish to fix it.
For the construction to be topos-internal, $x_0$ must be a \emph{global element} $x_0 : \mathbf{1} \Rightarrow X$, that is, a value $x_0\,c \in X(c)$ at each context, natural in $c$: $X(f)(x_0\,c) = x_0\,d$ for every $f : d \to c$.
Naturality is what makes ``$X = x_0$'' a sub-presheaf, and hence classifiable.
A bare context-indexed family $c \mapsto x_0\,c$ that ignored restriction would not cut out a subobject, and the construction below would fail to be natural.

\paragraph{The classifier}
Define $\chi : X \Rightarrow \Om$ by sending $b \in X(c)$ to the sieve of arrows along which $b$ restricts to the fixed value:
\[ \chi_c(b) \;=\; \{\, f : d \to c \;\mid\; X(f)(b) = x_0\,d \,\}. \]
This is a sieve (closure under precomposition is naturality of $x_0$ together with functoriality of $X$), and $\chi$ is a mechanism: it commutes with restriction, again by functoriality. The Agda statement is the record below (where \texttt{pt d} is the global element $x_0$ evaluated at $d$); the proof obligations are the two equations just named.
\begin{lstlisting}
χ-sieve : (c : Ob) → F₀ X c → Sieve c
χ-sieve c b = (λ d f → (F₁ X f b ≡ pt d) , isSetF₀ X d _ _) , χ-closed

χ : Nat X Ω
χ = (λ c b → χ-sieve c b) , χ-nat
\end{lstlisting}

\paragraph{Classification}
The map $\chi$ classifies the subobject ``$X = x_0$'':
\[ \chi_c(b) = \top \quad\Longleftrightarrow\quad b = x_0\,c. \]
Both directions are short.
If $b = x_0\,c$ then for every $f : d \to c$ we have $X(f)(b) = X(f)(x_0\,c) = x_0\,d$ by naturality, so every arrow lies in $\chi_c(b)$ and the sieve is maximal.
Conversely, if $\chi_c(b)$ is maximal then in particular the identity lies in it, which says $X(\mathrm{id})(b) = x_0\,c$, i.e.\ $b = x_0\,c$.
Specialising the first direction to $b = x_0\,c$ gives that the value the intervention forces is always classified true: $\chi_c(x_0\,c) = \top$.

\paragraph{What $\chi$ names, and what it does not}
This $\chi$ is the characteristic map of the target the companion paper's intervention forces, not a parallel construction.
The internal $\dom{X}{x_0}$ sets the prior at context $c$ to the point mass at $x_0\,c$, whose support is the singleton $\{x_0\,c\}$.
Two lemmas tie this to $\chi$: \texttt{do-prior} shows the intervened prior is exactly that point mass, and \texttt{do-classified} shows $x_0\,c$ is the point $\chi$ classifies true.
So $\chi$ is the topos-internal name of the value the intervention pins $X$ to.
It is not, by itself, the name of the \emph{operation}: at the level of $\Om$ the subobject ``$X = x_0$'' is the same whether it is reached by intervening or by observing, and $\chi$ classifies that target without telling the two apart.
The distinction is a property of the mechanisms beneath $\Om$, which the next section establishes.

\paragraph{Why the target, and not the sub-model}
The source states the intervention differently: a sub-model of an SCM $M$ ``is simply a monic arrow $f_x : M_x \hookrightarrow M$''~\cite{mahadevan-2025a,mahadevan-2025c}, with Pearl's $M_x = \langle U, V, F_x \rangle$, $F_x = \{f_i : V_i \notin X\} \cup \{X = x\}$, as the intended subobject.
The arrow is never instantiated. The general shape is given --- a commuting square whose two vertical maps $i$ and $j$ are stipulated monic --- and the source asserts that ``Definition~14 \ldots\ in fact explicitly defines one such (monic) arrow between the model $M$ and its submodel $M_x$''~\cite{mahadevan-2025c}. But Definition~14 is Pearl's submodel definition verbatim, and defines no arrow: no $i$, no $j$, no monicity.
Read in the setting the source's own classifier fixes --- objects are functions $f : U \to V$, and $\Om$ is the three-element $\{0,\tfrac{1}{2},1\} \to \{0,1\}$, the classifier of the arrow topos --- a monic $M_x \hookrightarrow M$ is a pair of injections $i : U \hookrightarrow U$ and $j : V \hookrightarrow V$ with $F \circ i = j \circ F_x$, and it forks.
If $i$ and $j$ are identities, the square forces $F_x = F$, which excludes every non-trivial intervention.
If the sub-model is instead given smaller variable sets --- the realisations on which $X$ already takes the value $x$ --- the inclusion is monic, but what it describes is selection: conditioning, not surgery.
Pearl's $M_x$ is neither: it has the same exogenous and endogenous variables as $M$ and differs only in the mechanism, so it is not a subobject of $M$.

This is why we classify the value-fixing subobject and not the sub-model.
What $\Om$ can name is the target ``$X = x_0$'', and it names it uniquely.
What it cannot name is the difference between fixing that value by surgery and finding it by observation: the two determine the same subobject, so no map into $\Om$ separates them.
The next section shows that the difference is real, and that it lives one level down, among the kernels.

\paragraph{The classifier in general}
The intervention is one instance of a general fact we also prove: $\Om$ classifies sub-presheaves, which is the standard characterisation of the subobject classifier.
For \emph{any} restriction-closed predicate $P$ on a presheaf $B$ --- a sub-presheaf --- the map $\chi^P_c(b) = \{\, f : d \to c \mid B(f)(b) \in P \,\}$ is natural, its $\top$-fibre is exactly $P$ in both directions, and it is the \emph{unique} such map: any natural $\chi' : B \Rightarrow \Om$ whose $\top$-fibre is $P$ equals $\chi^P$.
Uniqueness turns on a single fact about sieves --- one containing the identity is maximal, by closure under precomposition --- which forces the membership of $\chi'_c(b)$ at each arrow $f$ to be ``$B(f)(b) \in P$'', the defining clause of $\chi^P$.
The value-fixing classifier above is the case $P = \{x_0\}$ (\texttt{Topos.Classifier}, with the intervention recovered as \texttt{do-classifier}).

\section{The do-operator is not conditioning}\label{sec:dosee}

The classifier of Section~\ref{sec:classifier} names the target an intervention forces --- the same subobject, we saw, whether that target is reached by $\dom{X}{x_0}$ or by conditioning on $X = x_0$.
What separates the two is causal, and it lives one level down, among the kernels: an intervention is surgery on the model, and surgery is not conditioning.
The inequality itself is classical --- it is the confounding Pearl's do-calculus is built to handle~\cite{pearl-causality}; what is new here is its machine-checked internalisation, beneath an $\Om$ that cannot tell the two apart.
We make this a theorem.

\paragraph{A confounder}
Take a common cause $U$ with edges $U \to X$ and $U \to Y$ and no direct $X \to Y$ edge --- a pure confounder.
The do-operator is graph surgery: $\dom{X}{x_0}$ severs the incoming edge $U \to X$ and pins $X$ to $x_0$, leaving $U$ and the $Y$-mechanism untouched.
Conditioning on $X = x_0$ does something else --- it reweights $U$, and through $U$ the outcome $Y$.
On the binary model with $U \sim \mathrm{Bernoulli}(p)$ for any $0 < p < 1$, $X = U$, and $Y = U$ --- perfect confounding --- the interventional and observational answers differ:
\[ \Pr(Y = 1 \mid \dom{X}{1}) = p, \qquad \Pr(Y = 1 \mid X = 1) = 1, \]
where $p = \Pr(U{=}1)$ is the prior on the common cause.
Intervening on $X$ cannot touch $Y$, whose law stays $\Pr(U)$; observing $X = 1$ forces $U = 1$ and with it $Y = 1$.

\paragraph{The theorem}
Both sides are computed in the companion paper's probability monad.
The interventional marginal is the pushforward of the surgically altered model; the observational side is the conditional-probability quotient $\Pr(X{=}1, Y{=}1) / \Pr(X{=}1)$, formed with the development's own partial division --- the definition of conditional probability, not a stand-in.
The two are distinct weights, $p \ne 1$, so
\[ \Pr(Y \mid \dom{X}{1}) \;\ne\; \Pr(Y \mid X = 1) \]
is a machine-checked inequality (module \texttt{DoSeeDistinct}).
It is not vacuous: we exhibit a concrete strictly-interior weight $\tfrac12$ (with $0 < \tfrac12 < 1$) at the rational layer, so a confounded model on which the two disagree exists rather than being merely posited.

Two facts frame the result.
On the do-side the outcome's law does not depend on the value forced on $X$ --- there is no $X \to Y$ edge to carry it --- so $\dom{X}{1}$ and $\dom{X}{0}$ give the same $Y$-marginal (\texttt{do-joint-Y-invariant}: the $Y$-marginal of the intervened joint reduces, by the monad laws, to $\Pr(U)$ bound through $k_Y$, which never mentions the forced value); observation has no such invariance.
And the surgery pins $X$ exactly where the classifier looks: under $\dom{X}{x_0}$ the $X$-marginal is the point mass at $x_0$ (\texttt{do-fixes-X}), the value $\chi$ of Section~\ref{sec:classifier} classifies true.
The classifier gives the intervention its internal name; this section gives it its content.

\section{Collating local mechanisms: a pullback}\label{sec:gluing}

When two mechanisms are computed over a shared target, they combine into a single one: their pullback over that target.
Mahadevan motivates sheaves with a stronger claim --- that ``local functions can be collated together to yield a unique global function''~\cite[\S6]{mahadevan-2025c} --- but proves no existence theorem for collation; the local-to-global result of the companion~\cite{mahadevan-2025b} assumes a sheaf of causal kernels and establishes only uniqueness.
Collation over a \emph{cover} is the sheaf condition, which a presheaf need not satisfy once a topology is fixed; the canonical way to supply it is sheafification, which we do not formalise and whose effect on the distribution carrier we leave open (Section~\ref{sec:scope}).
What a presheaf topos does provide is the limit: two mechanisms agreeing on a common target have a pullback, computed contextwise, with a universal property, and that pullback is what we verify here.
It always exists. The interesting case is therefore not the pullback but assembly over a cover: over three contexts, pairwise-agreeing local data need not assemble into a global section at all (Section~\ref{sec:obstruction}).

\paragraph{The pullback}
Let $p : X \Rightarrow Z$ and $q : Y \Rightarrow Z$ be two mechanisms with a common target $Z$, against which agreement is measured.
Their pullback $X \times_Z Y$ has, at each context $c$, the pairs that agree in $Z$:
\[ (X \times_Z Y)(c) \;=\; \{\, (x,y) \in X(c)\times Y(c) \;\mid\; p_c(x) = q_c(y) \,\}. \]
The agreement condition is a proposition (equality in the set $Z(c)$), so the fibre is a set and restriction preserves it: restricting an agreeing pair gives an agreeing pair, using naturality of $p$ and $q$.
This makes $X \times_Z Y$ a presheaf with two projections $\pi_1, \pi_2$ to $X$ and $Y$, and the square $p \circ \pi_1 = q \circ \pi_2$ commutes by construction: the commuting witness is just the stored agreement.

\paragraph{The universal property}
A \emph{cone} is a presheaf $A$ with two mechanisms $a : A \Rightarrow X$, $b : A \Rightarrow Y$ that agree on $Z$, i.e.\ $p \circ a = q \circ b$ pointwise.
It collates to a \emph{unique} global mechanism $\langle a, b\rangle : A \Rightarrow X \times_Z Y$:
\begin{lstlisting}
glue      : Nat A Pullback
glue-π₁   : (c)(z) → fst π₁ c (fst glue c z) ≡ fst a c z
glue-π₂   : (c)(z) → fst π₂ c (fst glue c z) ≡ fst b c z
glue-uniq : (u : Nat A Pullback)
          → (π₁ ∘ u ≡ a) → (π₂ ∘ u ≡ b) → u ≡ glue
\end{lstlisting}
Existence is immediate: send $z$ to the agreeing pair $(a_c(z), b_c(z))$, whose agreement is the cone's hypothesis; naturality is naturality of $a$ and $b$ on the two components, with the agreement coherence automatic because it is a proposition.
The projection equations hold by definition.
Uniqueness is the only step with content: a competing mechanism $u$ whose projections are $a$ and $b$ must equal $\langle a,b\rangle$, because a pair is determined by its two components and the agreement witness is again propositional.
This is the pullback form of collation --- two mechanisms agreeing on a target combine into a unique third --- a limit in the presheaf topos, needing neither a topology nor sheafification.

\section{The internal language}\label{sec:forcing}

A \emph{predicate} on a presheaf $B$ is a mechanism $\varphi : B \Rightarrow \Om$, i.e.\ a $B$-indexed family of truth values, equivalently a subobject of $B$.
Kripke-Joyal semantics interprets the connectives of the internal language as operations on $\Om$ and determines, for each context $c$ and each generalised element $a \in B(c)$, whether $c$ \emph{forces} $\varphi$ at $a$.

\paragraph{Forcing}
We define forcing of a truth value $S$ at a context $c$ by ``the identity of $c$ lies in $S$'':
\[ c \Vdash S \;:=\; (\mathrm{id}_c \in S). \]
Since a sieve containing the identity is the maximal sieve, this agrees with ``$S = \top$''; the identity formulation is prop-valued and keeps the clauses at the value level.
Forcing a predicate at an element is forcing its truth value: $c \Vdash_\varphi a := c \Vdash \varphi_c(a)$.

\paragraph{Locality}
The defining property of a Kripke semantics is that forcing is stable under restriction: if $c \Vdash S$ and $f : d \to c$, then $d \Vdash S{\cdot}f$ (the pullback of $S$).
This holds because a sieve is downward closed.
At the level of predicates it says: if $a$ satisfies $\varphi$ at $c$, so does every restriction of $a$; this internal-logic form is proved from naturality of $\varphi$.

\paragraph{The clauses}
We verify the Mac Lane-Moerdijk clauses~\cite[VI.6]{maclane-moerdijk-1992} specialised to a presheaf topos.
Conjunction and disjunction are local at the context, as in any Kripke model:
\[ c \Vdash S \wedge T \iff (c \Vdash S)\text{ and }(c \Vdash T), \qquad
   c \Vdash S \vee T \iff \lVert (c \Vdash S) + (c \Vdash T) \rVert. \]
Both hold definitionally once the sieve operations are pointwise meet and join; the join carries a propositional truncation, which is why disjunction is recorded up to $\lVert-\rVert$.
Implication is the characteristic Kripke-Joyal clause, quantifying over all future contexts:
\[ c \Vdash S \Rightarrow T \iff \text{for all } f : d \to c,\ (d \Vdash S{\cdot}f)\text{ implies }(d \Vdash T{\cdot}f). \]
Falsity is forced nowhere, $c \Vdash \bot$ is empty.
For the quantifiers, over a value presheaf $D$, a predicate $\varphi$ on $B \times D$ yields predicates $\forall_D\varphi$ and $\exists_D\varphi$ on $B$, with clauses
\[ c \Vdash \forall_D\varphi\,(a) \iff \forall (f : d\to c)\,(t \in D(d)),\ d \Vdash_\varphi (a{\cdot}f,\,t), \]
\[ c \Vdash \exists_D\varphi\,(a) \iff \big\lVert\, \textstyle\sum_{t\in D(c)}\ c \Vdash_\varphi (a,\,t) \,\big\rVert. \]
The universal quantifier ranges over all restrictions and all elements; the existential is local-existential at the context (the image of the projection), in the presheaf-Kripke style.
Each quantifier object carries full sieve structure (downward closure and naturality), which we verify; most of the formalisation effort lies there rather than in the clauses themselves.
The source states these clauses for a general topos and refers their proof to Mac Lane-Moerdijk~\cite[Theorem~12]{mahadevan-2025a}; what we add is their verification for the presheaf case --- each binary connective and both quantifiers in both directions, together with the $\top$ and $\bot$ clauses and the locality property --- carried out in a proof assistant.

\section{A Lawvere-Tierney do-calculus}\label{sec:modal}

The modal layer interprets a \emph{Lawvere-Tierney topology} $j : \Om \to \Om$ as a causal modality: $j\,S$ is the closure of $S$, and the $j$-closed truth values are those stable under the localisation $j$ names.

\paragraph{Inflationarity is derivable}
A Lawvere-Tierney topology is usually given by three equations: $j\top = \top$, $j(S \wedge T) = j\,S \wedge j\,T$, and $j(j\,S) = j\,S$ (Mahadevan's Definition~10~\cite{mahadevan-2025b} states exactly these).
A closure operator also needs the unit $S \le j\,S$ (inflationarity), and it is tempting to add it as a fourth axiom. On the subobject classifier one need not: a topology is a \emph{natural} morphism $j : \Om \Rightarrow \Om$, and naturality together with $j\top = \top$ already force inflationarity. For $f \in S$ the restriction $S{\cdot}f$ is the maximal sieve (sieves are downward closed), so by naturality $(j\,S){\cdot}f = j(S{\cdot}f) = j\top = \top$, whence $f \in j\,S$; we machine-check this derivation (module \texttt{InflationarityDerivable}).
What the three equations do \emph{not} force is inflationarity for an operator on an \emph{arbitrary} Heyting algebra, where $j$ is a bare monotone map and naturality has no meaning: a three-element chain $\bot < a < \top$ with $j(a) = \bot$ satisfies all three yet collapses $a$ below itself (module \texttt{InflationarityIndependence}).
The naturality of $j$ on $\Om$ is thus what makes it a closure operator (a \emph{nucleus}): a formalisation that models topologies on a bare Heyting algebra must add inflationarity by hand, but on the classifier it is a theorem.
\begin{lstlisting}
record LawvereTierney where field
  jop    : (c) → Sieve c → Sieve c
  jnat   : IsNat Ω Ω jop          -- j is a mechanism Ω ⇒ Ω
  j-⊤    : jop c ⊤ ≡ ⊤
  j-∧    : jop c (S ∧ T) ≡ jop c S ∧ jop c T
  j-idem : jop c (jop c S) ≡ jop c S
-- inflationarity  f ∈ S → f ∈ jop c S  is DERIVED (jnat + j-⊤)
\end{lstlisting}

\paragraph{The modality as a reflector}
Writing $\cl = j$, the data above make $\cl$ a \emph{reflective} modality on truth values --- one that sends each truth value to the least $j$-closed value above it (its reflector). The unit $\eta : S \le \cl S$ is inflationarity, $\cl S$ is modal ($j$-closed) by idempotence, and $\cl$ is monotone, which we derive from meet-preservation rather than assume.
The reflector universal property follows: for a modal $T$,
\[ S \le T \quad\Longleftrightarrow\quad \cl S \le T. \]
So the $j$-closed truth values form a reflective sub-poset of $\Om$, with $\cl$ the reflector; this is the propositional shadow of sheafification.

\paragraph{A concrete topology}
The identity is a topology but a vacuous one.
The source never specialises $j$ on $\Om$ to a concrete operator, its instances being Grothendieck covers of regimes~\cite{mahadevan-2025b}; we provide the double-negation topology $\nn$ as a non-degenerate instance, for which the $j$-closed truth values are the classically-behaved ones.
This needs the Heyting structure on sieves --- bottom, implication, negation --- which we build, and then the three topology axioms plus naturality for $\nn$.
Two are short (truth-preservation, and idempotence via the triple-negation law $\nn\neg = \neg$); meet-preservation is the substantial one, whose difficult direction $\nn S \wedge \nn T \le \nn(S \wedge T)$ is the intuitionistically-valid but non-trivial step.
With $\nn$ established, the modal results below are no longer vacuous.

\paragraph{Interventions are modal}
The upshot is that interventions and the do-calculus survive localisation.
The truth value classifying the intervened value is $j$-closed for every topology:
\[ \text{$\chi_c(x_0\,c)$ is $j$-closed,} \]
because it equals $\top$ (Section~\ref{sec:classifier}) and $\top$ is $j$-closed for every topology.
Likewise each of Pearl's three rules, as verified in the companion paper, has a conclusion that is an equality of distributions; that equality is a proposition (distributions form a set), so it has an internal truth value, and that truth value is $j$-closed for every topology.
\begin{lstlisting}
do-j-stable : (J)(c) → is-j-closed J c (χ-sieve c (pt c))
modal-rule1 : (J)(c) → is-j-closed J c (rule1-Ω c)
modal-rule2 : (J)(c) → is-j-closed J c (rule2-Ω c)
modal-rule3 : (J)(c) → is-j-closed J c (rule3-Ω c)
\end{lstlisting}
In causal terms, a do-fact and each do-calculus rule hold in the internal logic of every sheaf subtopos; they are invariant under any modality $\cl$.
At the double-negation topology this is concrete: the intervention sieve and Rule~1 are $\nn$-closed (module \texttt{NonTrivialModal}), the Boolean-localisation instance of the generic statement.

\paragraph{What this does and does not show}
These closure facts are true, but their proof carries no causal content, and we do not overstate them.
Each rule's conclusion is a proved equality of distributions, hence a proposition with internal truth value $\top$, and $j\top = \top$ is a topology axiom; the same holds for $\chi_c(x_0\,c)$, which equals $\top$ by Section~\ref{sec:classifier}.
Thus $\mathtt{modal\text{-}rule_i}$ would typecheck verbatim with the causal content deleted: the proof depends only on the conclusion's truth value being $\top$, not on the rule that produced it.
Mahadevan's thesis, however, is that $j$ \emph{selects} the causal claims stable under localisation~\cite{mahadevan-2025b}, which presupposes that some are not.
The substantive statement therefore concerns \emph{contingent} causal claims --- conditional independences, which hold at some regimes and fail at others --- whose truth value is a non-maximal sieve.

\paragraph{A contingent claim, and its stability}
We exhibit one (module \texttt{ContingentCI}).
Over two regimes, with values in a two-point type and rational kernels, take the claim that the $Y$-marginal is invariant under $\dom{X}{1}$ --- the very proposition $\mathtt{rule1\text{-}\Omega}$ internalises, now \emph{without} assuming $Y \perp X$.
At the regime whose mechanism is the constant kernel it holds, so its truth value is $\top$; at the regime whose mechanism is the copy kernel $Y := X$, the intervention shifts the marginal from the point mass at $0$ to the point mass at $1$, so it \emph{fails}.
The failure is a computation in the rational weight algebra, not a postulate: the two marginals are distinct point masses, separated by their mass at $1$ ($\mathtt{w1} \neq \mathtt{w0}$), giving
\[ \mathtt{witness\text{-}non\text{-}maximal} : \Sigma\,(c : \text{regime})\;\neg\,(\mathtt{ci\text{-}\Omega}\,c \equiv \top). \]
It gives the modal layer a claim not already $\top$ --- the case the modal-rule stability theorems could not reach.
The same non-maximal claim, taken through the classifier of Section~\ref{sec:classifier} rather than regime-by-regime, is a subobject of the terminal presheaf, i.e.\ an internal truth value $\mathbf{1} \Rightarrow \Om$ (module \texttt{CIObject}) --- the subobject shape Section~\ref{sec:transport} gives counterfactuals.
On it the stability question has content, and we answer it (module \texttt{ModalCI}): for the double-negation topology, $\mathtt{ci\text{-}\Omega}$ is $j$-closed at \emph{both} regimes --- at one because its value is $\top$, at the other because its value is $\bot$ (the empty sieve, which is $\nn$-closed), and $\bot$ is not $\top$.
The verdict of a contingent independence, whether it holds or fails, therefore survives the Boolean localisation --- $j$-stability applied to something other than $\top$.
This escapes the $\top$-vacuity above but not degeneracy in general: the base here is the \emph{discrete} two-regime category, on which $\Om$ is Boolean and $\nn$ is the identity, so every truth value is $\nn$-closed automatically. The non-degenerate case is the intervention coverage below.
This is the $j$-do-calculus claim in the case where the statement is not already $\top$. The general soundness that would let $\mathtt{ci\text{-}\Omega}$ range over arbitrary graphical independences (Section~\ref{sec:scope}) remains future work, and not by routine extension. The sieve structure holds only where refinement preserves the independence, and for a collider --- a variable with two arrows into it, $X \to C \leftarrow Y$ --- it does not: conditioning on the collision vertex $C$ opens the path between its causes (Berkson's paradox). That failure of restriction-stability is the obstruction the general case must address --- \texttt{CIObject} builds the contingent-independence subobject and records the collider as that scope boundary, in comments, not as a theorem.

\paragraph{The modality does causal selection}
The stability above is limited. A \emph{site} is a category of contexts together with a notion of which arrow-families \emph{cover} an object; on the two-regime \emph{discrete} site the only covering families are trivial, so $j$ never \emph{collates} across contexts, and the $\top$-value of $\mathtt{modal\text{-}rule}_i$ is a degeneracy of that coverage rather than a feature of the framework.
The selection Mahadevan's thesis wants --- $j$ closing a claim that holds at every context \emph{below} up to holding at the one above --- needs a site with a non-trivial coverage.
The companion $j$-do-calculus~\cite{mahadevan-2025b} specifies such coverages on paper, as observational and interventional chart bases over a graph.
We give a minimal causal base category and compute in it (modules \texttt{InterventionSite}, \texttt{InterventionModality}): the intervention poset $\mathtt{do}_0 \to \mathtt{obs} \leftarrow \mathtt{do}_1$, with the observational context $\mathtt{obs}$ intended to be covered by its two interventions $\{\mathtt{do}_0, \mathtt{do}_1\}$ --- a non-maximal sieve, impossible on the discrete site.
The induced closure is discriminating: a claim holding under \emph{both} interventions is covered up to holding observationally, while one holding under only \emph{one} survives as a proper sieve,
\[ j\,\{\mathtt{do}_0,\mathtt{do}_1\} \equiv \top, \qquad \neg\,(\,j\,\{\mathtt{do}_0\} \equiv \top\,). \]
On this intended coverage the closure at $\mathtt{obs}$ is thus non-degenerate --- the two decisive memberships are what we machine-check. We formalise neither the coverage as a Grothendieck topology nor the closure as a natural map $\Om \Rightarrow \Om$, and we do not verify the Lawvere-Tierney axioms for it: the operator is defined at $\mathtt{obs}$ and computed there. What the computation realises is the invariance-across-interventions selection the $j$-do-calculus intends; the $\top$-collapse is a property of the coverage, not an obstruction in the theory.
The directed / $\infty$-categorical lift (Section~\ref{sec:scope}) would let this range over the full space of interventions; the propositional, single-variable case is what we machine-check here.

\section{An obstruction: causal contextuality}\label{sec:obstruction}

The pullback of Section~\ref{sec:gluing} always exists, being a limit; but collation over a cover is a different matter, and it can fail.
Over a cover by three measurement contexts, a compatible family of local models --- one per context, agreeing on every shared marginal --- can have a support that admits no global section.
This is \emph{strong contextuality} in the sense of Abramsky and Brandenburger~\cite{abramsky-brandenburger-2011}, and in causal terms it is the statement that pairwise-consistent local causal data need not come from any single global model.
At least three contexts are needed, and the reason is combinatorial rather than a fact about limits: over a cover by two contexts, any local model in the support at one extends across the overlap, by compatibility, to one at the other, and the two together are a global section.
For the modular, ``map-reduce'' assembly of mechanisms from local data that motivates the gluing, this is a structural warning. Local pieces that agree wherever they overlap can still fail to assemble, so a procedure that stitches local causal models cannot certify a global one from pairwise agreement alone.
The witness below is minimal and built by hand, not drawn from data; it shows the failure mode occurs and is exactly located, not that it is common.

\paragraph{Specker's triangle}
The minimal witness has three Boolean observables $A, B, C$, measured pairwise, each pair perfectly anti-correlated: the context $\{A,B\}$ supports only $A \neq B$, and likewise $\{B,C\}$ and $\{A,C\}$.
Each context is locally realisable, and the family is \emph{compatible} (no signalling): every value of a shared observable occurs in both contexts that contain it, so the single-observable marginals agree.
But there is no global assignment, because $A \neq B$ and $B \neq C$ force $A = C$, against $A \neq C$.
We prove its three obligations: local realisability and the absence of a global section (bundled as a record), and compatibility, i.e.\ no-signalling (a separate lemma); the no-global obligation is discharged by deciding the eight Boolean assignments.
\begin{lstlisting}
no-global : ¬ GlobalSection
no-global ((true , true , _) , p , _ , _) = p refl
no-global ((true , false , true) , _ , _ , r) = r refl
...                                      -- six cases in all
\end{lstlisting}

\paragraph{A holonomy criterion}
The obstruction can also be computed, in the spirit of Abramsky, Mansfield and Barbosa~\cite{abramsky-mansfield-barbosa-2012}, and the computation generalises well beyond this example.
Over an \emph{arbitrary} set of observables and \emph{arbitrary} abelian coefficients, we define the coboundary $\delta$ and the holonomy of a closed walk --- the sum of a $1$-cochain along it --- and prove the telescope: a coboundary's holonomy is the difference of its endpoints, hence trivial around a loop.
It follows that any $1$-cochain with non-trivial holonomy around a closed walk is not a coboundary on that walk's edges.
The causal reading is direct. A candidate global model is a labelling of the observables, a $0$-cochain, whose coboundary should realise the observed pairwise correlations, a $1$-cochain on the measured pairs; so a correlation cochain that is not a coboundary is one that no global labelling explains.

The criterion discriminates, which is what makes it more than a second proof of the same fact.
Specker's model --- the cochain anti-correlated on every edge of the triangle --- has holonomy $1$, so it is not a coboundary, and the obstruction is a computed, non-trivial class.
A four-cycle is the contrasting instance: its holonomy is trivial, and the even cycle is moreover \emph{satisfiable}, as we machine-check a global two-colouring that is a coboundary on all four of its edges (\texttt{square-coboundary}) --- the global section the triangle lacks.
For the triangle we also machine-check both inclusions $\operatorname{im}\delta = \ker(\mathrm{hol})$ --- every edge-coboundary has zero holonomy, and every zero-holonomy cochain is an edge-coboundary --- together with surjectivity of holonomy; $H^1 \cong \mathbb{Z}_2$ follows, though we do not formalise the quotient itself.
And because a global model is exactly an edge-coboundary witness, the class recovers the possibilistic no-global-section fact proved above: two formalisations of one obstruction, agreeing.

What we machine-check is the degree-one, constant-coefficient fragment of that line. The full invariant of Abramsky, Mansfield and Barbosa takes its class in the \v{C}ech cohomology of the nerve of the cover, with coefficients in an abelian presheaf derived from the support, and is sufficient but not necessary --- they exhibit a strongly contextual model whose class vanishes.
Our complex is not that nerve: we take the observables as vertices and the measured pairs as edges. (For the triangle the nerve is also a $3$-cycle, so the two agree here; in general they do not.)
Those presheaf coefficients, the higher cochain degrees, and the nerve of an arbitrary cover are the remaining generalisation.

\section{Counterfactual transport as an invariance modality}\label{sec:transport}

The modal layer of Section~\ref{sec:modal} shows that interventions and Pearl's rules are $j$-stable for every topology: invariant under the localisations a Lawvere-Tierney topology names.
A natural question, which Section~\ref{sec:scope} flags, is whether the internal logic yields a \emph{transportability} criterion --- whether a causal effect established in one domain can be transferred to another, where it need not take the same value --- of the kind Bareinboim and Pearl give graphically~\cite{bareinboim-pearl-2016}.
We answer its invariance case: across a cover of regimes, the transport of a counterfactual --- its holding at the global regime, and so, by locality, throughout the cover --- coincides with $j$-stability.
This is invariance across regimes, not the full graphical transportability of Bareinboim and Pearl, a boundary Section~\ref{sec:scope} draws.
The development is in the same Cubical Agda, typechecks under \texttt{--safe}, and reuses the classifier, the forcing layer, and the modal layer verified above.

\paragraph{Counterfactuals as subobjects}
Fix a base category of regimes (environments) and a presheaf $W$ of worlds over it, a world carrying the exogenous data against which a counterfactual is evaluated.
A counterfactual query is a predicate on worlds that is \emph{restriction-stable}: if it holds of a world it holds of every restriction.
Restriction-stability is exactly sieve closure, so such a predicate cuts out a subobject, named by a characteristic map $\chi : W \Rightarrow \Om$ whose naturality we prove.
Forcing $\chi$ at a regime is the counterfactual holding there; we call this transport to that regime.
Kripke locality then yields transport-invariance immediately: a counterfactual forced at a regime is forced at every regime restricting into it.

\paragraph{Direct transportability is $j$-stability}
The identification is as follows.
A counterfactual transports to the global regime iff it is forced there, iff its internal truth value is the maximal sieve $\top$; and being $\top$ entails being $j$-closed for \emph{every} Lawvere-Tierney topology, so direct transportability entails invariance (the converse is the soundness result below, under a density hypothesis).
The middle step is restriction-stability carrying global truth down to the cover; the last is the truth-preservation axiom $j\top = \top$ --- the same $\top$-collapse by which Section~\ref{sec:modal} proves do-facts and Pearl's rules $j$-stable.
So the invariance modality the paper studies for interventions is, for counterfactuals, transport to the global regime:
\[ \text{holds at the global regime} \iff \text{$j$-stable (invariant)}, \]
the backward direction under the density and local-truth hypotheses of the soundness paragraph below.
A counterfactual that holds only in an environment, not globally, correspondingly fails to be $j$-closed for a topology under which the global regime is covered by its environments: forced at the environment but not at the global regime, exactly the failure of transport.
The qualification is needed --- under the identity topology every truth value is $j$-closed --- and this failure case we do not machine-check.

\paragraph{The probabilistic case}
The identification is not special to deterministic outcomes.
The finite-distribution monad $\mathtt{FDist}$ is a set, so a distributional statement is a proposition and a probabilistic counterfactual is again an internal truth value; the subobject and its forcing transfer unchanged. We carry this out for a point-mass outcome; lifting it to the companion paper's full interventional distribution, with abduction as Bayesian conditioning, is the natural next step and is not yet formalised.

\paragraph{Soundness}
The converse direction is transport \emph{soundness}: invariance entails direct transportability.
We prove that for any topology in which the environment is $j$-dense --- the identity lies in the closure of its sieve --- a $j$-stable counterfactual that holds in the environment holds globally.
In the present finite regime category the double-negation topology is dense in this sense: the negation of a counterfactual holding along the environment arrow is empty, so its double-negation closure is $\top$, which is the density hypothesis up to the identification of a maximal sieve with membership of the identity; density in a general base category is not claimed.
This is soundness in the shape of the companion paper's d-separation soundness, and of Bareinboim and Pearl's transport soundness, both of which carry admissibility side-conditions.
The matching completeness statement, an equivalence with their s-hedge criterion~\cite{bareinboim-pearl-2013}, we leave open; so too the discharge of the density side-condition into a single unconditional term, which the present finite regime category does not support cleanly.

\paragraph{Scope of this development}
This is the deterministic and finite-regime account: the worlds are a presheaf of exogenous data, the cover is small, and transport is the internal-logic notion just defined, not yet proved equivalent to the complete graphical criterion.
The $\top$-collapse and the $j$-closedness that identify transport with invariance are proved for an arbitrary restriction-stable regime predicate, but over this fixed two-regime cover; the forcing and classifier layers beneath them are general in the base category as well.
Neither half of the framing is new on its own. Counterfactual logic inside a topos, with a neighbourhood system of possible worlds, is de Araujo Fernandes and Haeusler's~\cite{fernandes-haeusler-2009}, and the source sketches counterfactuals over causal models on exactly that basis~\cite[\S9.2]{mahadevan-2025c}, following Lewis rather than Kripke-Joyal.
What we add is the modal identification and its proof: a counterfactual is a restriction-stable predicate, hence a subobject; forcing it at a regime is transport there; and transport to the global regime coincides with $j$-stability for every topology --- machine-checked, for an arbitrary such predicate.

\section{The formalisation}\label{sec:formalisation}

The core development is thirty-two Cubical Agda modules, with eight more for counterfactual transport (Section~\ref{sec:transport}), over an eight-module companion probability layer.
We indicate where the cubical setting is used essentially, which proofs were substantial, and two points that mechanisation forced.

\paragraph{Structure and reuse}
The base modules fix a category of contexts and the presheaves over it (\texttt{Cat}, \texttt{PSh}), build the classifier of sieves $\Om$ (\texttt{Omega}), and re-export the companion paper's finite-distribution monad as an internal distribution object $\mathrm{Dist}\,X = \mathtt{FDist}\circ X$ (\texttt{InternalDist}).
On top sit the pillars --- the intervention classifier (\texttt{DoClassifier}, with $\Om$'s general universal property and uniqueness in \texttt{Classifier}), the do-operator's separation from conditioning (\texttt{DoSeeDistinct}), collation by pullback (\texttt{Gluing}), forcing (\texttt{Forcing}), the modal layer (\texttt{LawvereTierney}, \texttt{InflationarityDerivable}, \texttt{Modality}, \texttt{DoubleNegation}, with the contingent-independence stability of \texttt{ContingentCI} and \texttt{ModalCI} and the intervention coverage of \texttt{InterventionSite} and \texttt{InterventionModality}), and the obstruction (\texttt{Contextuality}, \texttt{Cohomology}, \texttt{CechCohomology}) --- together with the internalised do-calculus rules (\texttt{Rule1}--\texttt{Rule3}, \texttt{ModalRules}) and the counterfactual-transport development (the \texttt{Transport} modules).
The mechanisms themselves are not rebuilt: the internal kernels are the companion paper's, and the topos layer only organises them.
The Agda development assumes no axioms (the directed \texttt{rzk} companion of Section~\ref{sec:scope} assumes one, extension extensionality).
The probability monad is built over an ordered field that earlier drafts left as a \texttt{postulate}; we now supply that field concretely as the cubical-library rationals $\mathbb{Q}$, each of its ring, order, and bounded-division laws a proved theorem.
One subtlety makes this more than a substitution: a transparent $\mathbb{Q}$ unfolds to stuck set-quotient projections, which defeats the unifier and breaks implicit-argument inference at hundreds of downstream sites --- the very reason the field had been left abstract.
We recover the abstract behaviour by re-exporting the field through a single \texttt{opaque} block, which keeps the operations rigid for unification while concretely equal to $\mathbb{Q}$; every downstream proof then goes through unchanged.
The whole development, probability layer and topos layer alike, typechecks under Agda's \texttt{--safe} flag --- which rules out postulates and any bypass of the termination, positivity, or universe checks --- and contains no unsolved holes.
The account is therefore not merely postulate-free relative to an assumed interface: it is unconditionally machine-checked, with the ordered field discharged at $\mathbb{Q}$.

\paragraph{What the cubical setting provides}
Sieve membership is valued in propositions, which keeps each $\Om(c)$ a set and lets us prove sieve equality from membership alone; the same device makes the agreement condition in the pullback and the commuting square of the gluing proof hold automatically, because a proof of a proposition is unique.
Two of the forcing clauses then hold \emph{definitionally}: $c \Vdash S \wedge T$ and the pair of $c\Vdash S$ and $c\Vdash T$ are literally the same term, and likewise for $\vee$ against the truncated sum, so those soundness proofs are the identity.
Propositional truncation $\lVert-\rVert$ carries disjunction, the local-existential quantifier, and the compatibility (no-signalling) witnesses of the contextuality model.
Function extensionality and the propositional-extensionality path $\Leftrightarrow$-to-$\equiv$ give the naturality and extensionality lemmas that every universal-property argument runs through.

\paragraph{The substantial proofs}
Three steps account for most of the difficulty.
First, the double-negation topology needs the Heyting structure on sieves built by hand --- bottom, implication, negation --- and its substance is the intuitionistically-valid meet inclusion $\nn S \wedge \nn T \le \nn(S\wedge T)$: the witness restricts both double-negations along the test arrow and reassembles an $(S\wedge T)$-membership from an $S$- and a $T$-witness, reassociating every composite by hand; idempotence then follows easily from the triple-negation law $\nn\neg = \neg$.
Second, the quantifier objects of \texttt{Forcing} are not the one-line clauses they reduce to: each of $\forall_D$ and $\exists_D$ must be given full sieve structure --- downward closure and naturality --- with the substitutions along associativity and functoriality discharged explicitly, and $\exists_D$ threaded through the truncation.
Third, the gluing universal property is a limit argument: existence is the agreeing pair, but uniqueness needs that a cone map is determined by its two components, with the agreement coherence automatic because it is propositional.

\paragraph{Two consequences of mechanisation}
Formalising forced one design choice and collapsed one family of proofs.
The intervention classifier typechecks only when $x_0$ is a \emph{natural} global element: a bare context-indexed family does not cut out a subobject and $\chi$ fails to be natural, so the naturality side-condition of Section~\ref{sec:classifier} is something the type system insisted on, not something we imposed.
And the modal-stability results --- that the intervention sieve and Pearl's three rules are $j$-closed for \emph{every} topology --- collapse to a single three-line argument: each conclusion is an equality of distributions, hence a truth value equal to $\top$, and $\top$ is fixed by every $j$.
The contextuality obstruction, by contrast, is decided concretely: the no-global-section lemma rules out all eight Boolean assignments, and the holonomy obstruction is a theorem over an arbitrary observable set and arbitrary abelian coefficients (\texttt{CechCohomology}); on the triangle's cover, with $\mathbb{Z}_2$ coefficients, the class is non-zero. The Specker triangle then has holonomy $1$ and is contextual, while a four-cycle has holonomy $0$ and, correctly, is not.

\section{Scope and limitations}\label{sec:scope}

We have formalised the 1-topos core: the classifier, the do-operator's separation from conditioning, gluing, the internal language, and the modal do-calculus, all over a presheaf topos.
Three things are deliberately outside it.

First, the modality acts on \emph{truth values}, not on arbitrary types.
A full type-level reflector (sheafification of an arbitrary presheaf, with its descent) is a larger construction --- the plus-construction, applied twice --- and needs no machinery beyond what Cubical Agda offers.
We have built the propositional reflector that the do-calculus actually uses.
Whether that larger construction leaves the distribution object intact is a further question, and we leave it open: our distributions are normalised contextwise ($\mathrm{Dist}\,X = \mathtt{FDist}\circ X$), so a section is already a quotient of weights, and the plus-construction's separating step quotients again, identifying any two sections that agree on some covering family.
That a functional reading such a section --- an adjustment weighting strata by a confounder's marginal, say --- still descends is not a consequence of the construction being standard.

Second, we work over presheaves, not sheaves on an instantiated site.
This matches the topos-causal-model papers~\cite{mahadevan-2025a,mahadevan-2025c}, whose constructions are presheaf constructions and whose Grothendieck topology is defined but never instantiated. The companion $j$-do-calculus~\cite{mahadevan-2025b} does instantiate one, giving observational and interventional chart bases on named collider graphs; we do not formalise those covers.
Pillars one (classifier) and three (internal logic) transfer to a sheaf topos unchanged; the gluing pillar is where covers would enter --- our pullback is the limit a presheaf topos already carries, whereas collation over a cover is the sheaf condition, which we do not supply and, for the reason just given, do not assume sheafification would supply.

Third, this is the \emph{undirected} account, and we can now say precisely what a directed one keeps and what it costs.
Causation is asymmetric, and the identity types of cubical type theory are symmetric: a causal arrow modelled as a path $x = y$ silently yields its inverse, collapsing ``$X$ causes $Y$'' into ``$Y$ causes $X$''.
A faithful treatment instead models causal influence as a \emph{non-invertible} arrow, and we have machine-checked the structural core of one.
This is a companion development in the directed cousin of cubical type theory --- the simplicial type theory of Riehl and Shulman~\cite{riehl-shulman-2017}, in which a type can carry a primitive notion of directed arrow that, unlike a path, has no inverse --- in the rzk proof assistant on the sHoTT library, verified under \texttt{rzk 0.8}, in the \texttt{directed/} directory of the artifact repository.

The translation is direct.
A causal world is a \emph{Segal type}, a type whose arrows compose; causal influence is one of those directed arrows, a \texttt{hom}, with no backward partner; transitivity of influence is composition of arrows.
A mechanism --- a downstream variable depending on an upstream one --- is a \emph{covariant family}, a family of types that transports forward along arrows, and interventional propagation is that forward transport.
On this dictionary we machine-check the structural laws of causal reasoning.
The trivial mechanism propagates a value unchanged.
A causal history propagates by composition, so influence through a mediator factors into its two legs --- associativity of causal composition (the one place the directed core invokes an axiom: the library's extension extensionality, the directed analogue of function extensionality).
A morphism between mechanisms commutes with propagation, so mechanisms transform naturally.
In a complete (\emph{Rezk}) world --- isomorphic variables already equal --- causal identity is causal isomorphism, by the completeness condition itself; and the Yoneda embedding realises a variable through the cone of its downstream effects. This is the shape of a Yoneda identifiability statement; we instantiate the embedding, not its full-faithfulness.
A counterfactual, finally, is the span underlying a twin network: two runs of the world sharing a single exogenous source.
The asymmetry is faithful throughout --- a forward \texttt{hom} yields nothing backward, where a symmetric path would have yielded the reverse automatically.

What this directed development cannot reach is exactly the do-operator.
An intervention $\dom{X}{x_0}$ mutilates the causal world, and to recover it, the directed account has to treat the universe of types as itself a directed world and extract the intervention as an arrow inside it.
That step is \emph{directed univalence}: the principle that an arrow between two types in the universe is the same data as a function between them, $\mathrm{hom}_{\mathcal{S}}(A,B) \simeq (A \to B)$.
It is the construction of Gratzer, Weinberger and Buchholtz~\cite{gwb-2024}, carried out in a modal extension of the theory (triangulated type theory) for $\mathcal{S}$, the universe of \emph{discrete} types --- the relevant regime here, since causal state spaces are sets. For the universe of categories the corresponding equivalence is modal, which is why the discrete case is the one to want.
rzk~0.8 provides part of that experimental substrate but disallows the amazing right adjoint the construction requires, and does not build the universe itself; and no \emph{synthetic} directed-univalent universe has been formalised in a proof assistant, to our knowledge and at the time of writing --- the one constructive model, Weaver and Licata's bicubical sets~\cite{weaver-licata-2020}, is semantic, with the cobar modality axiomatised. We can \emph{state} the goal in rzk but not inhabit it.
The two accounts thus meet at a precise boundary: the present, undirected one machine-checks the full 1-topos do-calculus and the invariance (direct-transportability) case, though its causal arrows are symmetric and invertible; the directed one keeps the asymmetry faithfully but stops short of the do-operator.
A machine-checked directed do-calculus, joining the two, requires a directed-univalent universe.

\paragraph{Invariance, not transportability}
The modality is easy to over-interpret, and warrants one caution.
A $j$-closed truth value is one that holds identically across the contexts $j$ glues: it is \emph{invariant}.
This is the ``globally valid'' interpretation the source intends for $j$~\cite{mahadevan-2025b}, and it is the natural home for invariant-prediction approaches to discovery, whose target is exactly the mechanism that does not change across environments.
It is \emph{not} transportability in the sense of Bareinboim and Pearl.
A causal effect can be identifiable in a target domain from source-experimental and target-observational data (hence transportable) while taking a different value in each domain, so it is not invariant and not $j$-closed.
Our modal-stability theorems (Section~\ref{sec:modal}) therefore certify invariance, the direct-transportability case, not transportability in general; the same caveat applies to the source, whose $j$ is the same invariance modality.
General transport is a marginalisation of a source factor against a target factor: the arithmetic is available (it is monadic bind in the underlying probability monad), and the selection structure separating shared from domain-specific mechanisms can plausibly be modelled by naturality, though we do not prove this.
Section~\ref{sec:transport} carries out this invariance case, identifying a counterfactual's invariance across a regime cover with $j$-stability and proving soundness; the completeness equivalence with the graphical s-hedge criterion~\cite{bareinboim-pearl-2013}, which is where effects taking a different value in each domain are decided, remains open.

\section{Related work}\label{sec:related}

The source of the framework is Mahadevan's topos causal models~\cite{mahadevan-2025a,mahadevan-2025c} and the intuitionistic $j$-do-calculus~\cite{mahadevan-2025b}; our development is a formalisation and, in the gluing and modal layers, a clarification of that work.
The categorical-probability background --- mechanisms as kernels, conditioning and disintegration as abstract theory --- is that of Markov categories~\cite{fritz-2020,cho-jacobs-2019}, and our internal mechanisms are instances of the probability monad verified in the companion paper~\cite{sargsyan-cubical-dsep}.
The do-operator itself has a categorical treatment in that line: Jacobs, Kissinger and Zanasi~\cite{jacobs-kissinger-zanasi-2019} represent an intervention as an endofunctor performing surgery on string diagrams, with the interventional distribution recovered through the interpretation functor. That is the same reading of intervention-as-surgery that Section~\ref{sec:dosee} confirms --- carried out on syntax, where the two operations are visibly different constructions, rather than on truth values, where they are not.
The topos-theoretic constructions follow Mac Lane and Moerdijk~\cite{maclane-moerdijk-1992}: the classifier of sieves, the Kripke-Joyal clauses, and Lawvere-Tierney topologies are all theirs.
The treatment of a topology as a reflective modality is the propositional case of the modalities of Rijke, Shulman and Spitters~\cite{rijke-shulman-spitters-2020}; the directed lift of the do-operator, which we leave to future work, draws on directed type theory.
The topos infrastructure we build on has been formalised before, though never for causal models: Quirin and Tabareau machine-check Lawvere-Tierney topologies and sheafification in homotopy type theory~\cite{quirin-tabareau-2016}, and the Cubical Agda library and its scheme-theoretic developments~\cite{zeuner-mortberg-2024} formalise presheaves and the sheaf condition. The universal property of the subobject classifier, which we verify here to ground the intervention and for self-containedness, is likewise standard and has been mechanised in general topos-theory libraries; we claim it as a completion, not a novelty. What is new here is the causal interpretation of this machinery --- not the machinery itself.

The book-length development of the framework~\cite{mahadevan-catagi-2026} includes a Lean~4 formalisation of its own. That development and ours do not overlap in substance. It is, by its author's own verification map, partial: the topos results record only a finite-limit core (the coalgebra topos is marked as future work), and the Grothendieck--Lawvere-Tierney correspondence is formalised in a single direction. It is classical, built on Mathlib, and does not aim at Agda's \texttt{--safe} discipline. It does not touch the intervention classifier's classification theorem, the gluing universal property, or the modal-stability results. Where it does address related material --- it states a Kripke-Joyal semantics declaration --- our contribution is not to be first to name the clauses but to prove all of them, constructively and axiom-free, for the causal setting.

A complementary line studies do-calculus \emph{derivations} combinatorially rather than semantically. Yvernes, Devijver, Clausel and Gaussier~\cite{yvernes-et-al-2026} organise the rewrites of Pearl's rules into a derivation graph, presupposing the soundness our companion paper mechanises, and show that do-calculus-equivalent estimands can differ in statistical efficiency. That distinction lives below the population layer any faithful semantics identifies, and is hence orthogonal to what a truth value in $\Om$ can see. Their open problem, extending the derivation graph with probabilistic rewritings that trivial identities render infinite, would be a sheafification: precisely the type-level modality that Section~\ref{sec:scope} defers to future work, along with the open question recorded there.

The obstruction of Section~\ref{sec:obstruction} belongs to a well-developed line that we do not claim to originate.
That contextuality is an obstruction to assembling a compatible family of local models into a global section is due to Abramsky and Brandenburger~\cite{abramsky-brandenburger-2011}; its witnessing by a \v{C}ech cohomology class --- sufficient for contextuality, though not necessary --- is due to Abramsky, Mansfield and Barbosa~\cite{abramsky-mansfield-barbosa-2012}. Both are in quantum foundations.
Gogioso and Pinzani~\cite{gogioso-pinzani-2023} carry the sheaf framework to causality and identify a ``causally-induced contextuality'', but in an operational setting covering arbitrary causal orders --- definite, dynamical or indefinite --- rather than for classical structural causal models.
For classical causal models the closest work is very recent: global counterfactuals can be obstructed by the homology of the causal graph~\cite{causal-counterfactual-obstructions-2026}, formalised there through cellular sheaves over optimal-transport spaces, a route different from ours and not machine-checked.
Quantum no-go inequalities have been machine-checked before --- the CHSH inequality and Tsirelson's bound in Isabelle/HOL~\cite{echenim-mhalla-mori-2023} and in Lean's Mathlib~\cite{mathlib-chsh} --- but those are bounds on correlations, not the global-section obstruction we treat here. Computer-verified Kochen-Specker results exist too~\cite{li-bright-ganesh-2024}, but as SAT and computer-algebra proof certificates rather than proof-assistant developments.
Against this background, our contribution is a machine-checked realisation of the obstruction (a compatible empirical model whose support admits no global section): to our knowledge it is the first such global-section obstruction verified in a proof assistant.
It is realised as finite combinatorics --- a possibilistic empirical model over three contexts --- and connects Mahadevan's $j$-stability programme, where obstructions do not appear, to the contextuality line.

\section{Conclusion}\label{sec:conclusion}

Topos causal models recast intervention, mechanism assembly and counterfactual reasoning as three topos-theoretic universal properties.
We have machine-checked the 1-topos core of this picture in Cubical Agda over a verified probability monad: the intervention classifier with its classification theorem (an instance of $\Om$'s universal property, which we also verify), the do-operator machine-checked to differ from conditioning on a confounder, the collation of local mechanisms as a pullback with its universal property, the Kripke-Joyal forcing clauses for every connective and quantifier, and a Lawvere-Tierney do-calculus in which interventions and Pearl's rules are stable under every topology.
Relative to the source, the development gives the pullback form of collation that the source only motivates, shows the subobject classifier names the target of an intervention but not the operation, which is surgery rather than conditioning, and shows the modal unit is derivable rather than a missing axiom --- inflationarity follows from naturality and $j\top = \top$; the double-negation topology makes the modal layer concrete where the source stays abstract.
The development is machine-checked under Agda's \texttt{--safe} flag, with the ordered field discharged concretely at $\mathbb{Q}$. The structural core of a directed account, in which the asymmetry of causation is primitive, we machine-check in rzk; the directed do-operator itself, which needs directed univalence, remains future work.

\section*{Acknowledgements}
I thank Jon Sterling for pointing out that inflationarity is derivable from $j\top = \top$ on the subobject classifier, which the modal layer of Section~\ref{sec:modal} now reflects, and for identifying further problems in an earlier version of this paper, which the present version corrects. Those observations are his; any errors that remain are mine alone.

\bibliographystyle{alphaurl}
\bibliography{references}

@book{pearl-causality,
  author    = {Judea Pearl},
  title     = {Causality: Models, Reasoning, and Inference},
  edition   = {2nd},
  publisher = {Cambridge University Press},
  year      = {2009},
}

@article{fritz-2020,
  author    = {Tobias Fritz},
  title     = {A Synthetic Approach to {M}arkov Kernels, Conditional Independence and Theorems on Sufficient Statistics},
  journal   = {Advances in Mathematics},
  volume    = {370},
  pages     = {107239},
  year      = {2020},
  doi       = {10.1016/j.aim.2020.107239},
}

@article{cchm-2018,
  author    = {Cyril Cohen and Thierry Coquand and Simon Huber and Anders M\"{o}rtberg},
  title     = {Cubical Type Theory: A Constructive Interpretation of the Univalence Axiom},
  journal   = {Journal of Applied Logics (IfCoLog)},
  volume    = {4},
  number    = {10},
  pages     = {3127--3170},
  year      = {2017},
  note      = {Conference version: TYPES 2015, LIPIcs 69, 5:1--5:34, 2018},
}

@inproceedings{cubical-agda,
  author    = {Andrea Vezzosi and Anders M\"{o}rtberg and Andreas Abel},
  title     = {Cubical {A}gda: A Dependently Typed Programming Language with Univalence and Higher Inductive Types},
  booktitle = {Proceedings of the {ACM} on Programming Languages ({ICFP})},
  volume    = {3},
  year      = {2019},
  pages     = {87:1--87:29},
  doi       = {10.1145/3341691},
}

@inproceedings{mahadevan-2025a,
  author    = {Sridhar Mahadevan},
  title     = {Universal Causal Inference in a Topos},
  booktitle = {Advances in Neural Information Processing Systems 38 ({N}eur{IPS} 2025)},
  year      = {2025},
}

@article{mahadevan-2025b,
  author  = {Sridhar Mahadevan},
  title   = {Intuitionistic $j$-Do-Calculus in Topos Causal Models},
  journal = {arXiv preprint arXiv:2510.17944v2},
  year    = {2025},
  note    = {Version~2, 25 July 2026; all references here are to that version},
}

@inproceedings{jacobs-kissinger-zanasi-2019,
  author    = {Bart Jacobs and Aleks Kissinger and Fabio Zanasi},
  title     = {Causal Inference by String Diagram Surgery},
  booktitle = {Foundations of Software Science and Computation Structures ({FoSSaCS} 2019)},
  series    = {Lecture Notes in Computer Science},
  volume    = {11425},
  publisher = {Springer},
  year      = {2019},
  doi       = {10.1007/978-3-030-17127-8_18},
}

@article{fernandes-haeusler-2009,
  author  = {Ricardo Queiroz de Araujo Fernandes and Edward Hermann Haeusler},
  title   = {A Topos-Theoretic Approach to Counterfactual Logic},
  journal = {Electronic Notes in Theoretical Computer Science},
  volume  = {256},
  pages   = {33--47},
  year    = {2009},
}

@article{echenim-mhalla-mori-2023,
  author  = {Mnacho Echenim and Mehdi Mhalla and Coralie Mori},
  title   = {A Formalization of the {CHSH} Inequality and {T}sirelson's Upper-Bound in {I}sabelle/{HOL}},
  journal = {Journal of Automated Reasoning},
  year    = {2023},
  note    = {Archive of Formal Proofs entry \texttt{TsirelsonBound}},
}

@misc{mathlib-chsh,
  author = {{The mathlib Community}},
  title  = {\texttt{Mathlib.Algebra.Star.CHSH}: {T}sirelson's inequality},
  year   = {2024},
  note   = {Lean~4 mathematical library},
}

@inproceedings{li-bright-ganesh-2024,
  author    = {Zhengyu Li and Curtis Bright and Vijay Ganesh},
  title     = {A {SAT} Solver and Computer Algebra Attack on the Minimum {K}ochen--{S}pecker Problem},
  booktitle = {Proceedings of the 33rd International Joint Conference on Artificial Intelligence ({IJCAI} 2024)},
  year      = {2024},
}

@book{mahadevan-catagi-2026,
  author    = {Sridhar Mahadevan},
  title     = {Categories for {AGI}},
  year      = {2026},
  publisher = {Draft manuscript},
  note      = {Lean~4 formalization companion at \url{https://github.com/sridharmahadevan/catagi}},
}

@inproceedings{yvernes-et-al-2026,
  author    = {Cl\'ement Yvernes and Emilie Devijver and Marianne Clausel and Eric Gaussier},
  title     = {Unveiling the Structure of Do-Calculus Reasoning via Derivation Graphs},
  booktitle = {Proceedings of the 43rd International Conference on Machine Learning ({ICML} 2026)},
  year      = {2026},
  url       = {https://arxiv.org/abs/2606.03719},
}

@article{cho-jacobs-2019,
  author  = {Kenta Cho and Bart Jacobs},
  title   = {Disintegration and {B}ayesian Inversion via String Diagrams},
  journal = {Mathematical Structures in Computer Science},
  volume  = {29},
  number  = {7},
  pages   = {938--971},
  year    = {2019},
  doi     = {10.1017/S0960129518000488},
}

@book{maclane-moerdijk-1992,
  author    = {Saunders {Mac Lane} and Ieke Moerdijk},
  title     = {Sheaves in Geometry and Logic: A First Introduction to Topos Theory},
  publisher = {Springer},
  year      = {1992},
  series    = {Universitext},
}

@misc{mahadevan-2025c,
  author = {Sridhar Mahadevan},
  title  = {Topos Causal Models},
  year   = {2025},
  eprint = {2508.08295},
  archivePrefix = {arXiv},
  primaryClass  = {cs.AI},
}

@article{rijke-shulman-spitters-2020,
  author  = {Egbert Rijke and Michael Shulman and Bas Spitters},
  title   = {Modalities in homotopy type theory},
  journal = {Logical Methods in Computer Science},
  volume  = {16},
  number  = {1},
  year    = {2020},
}

@misc{sargsyan-cubical-dsep,
  author        = {Karen Sargsyan},
  title         = {A cubical formalisation of conditional independence, {B}ayesian conditioning, and {P}earl's d-separation soundness},
  year          = {2026},
  eprint        = {2606.20351},
  archivePrefix = {arXiv},
  url           = {https://arxiv.org/abs/2606.20351},
  note          = {Companion paper},
}

@article{abramsky-brandenburger-2011,
  author  = {Samson Abramsky and Adam Brandenburger},
  title   = {The sheaf-theoretic structure of non-locality and contextuality},
  journal = {New Journal of Physics},
  volume  = {13},
  number  = {11},
  pages   = {113036},
  year    = {2011},
}

@misc{causal-counterfactual-obstructions-2026,
  author = {Rui Wu and Hong Xie and Yongjun Li},
  title  = {Cohomological Obstructions to Global Counterfactuals: A Sheaf-Theoretic Foundation for Generative Causal Models},
  year   = {2026},
  eprint = {2603.17384},
  archivePrefix = {arXiv},
  primaryClass  = {cs.LG},
}

@inproceedings{abramsky-mansfield-barbosa-2012,
  author    = {Samson Abramsky and Shane Mansfield and Rui Soares Barbosa},
  title     = {The Cohomology of Non-Locality and Contextuality},
  booktitle = {Proceedings 8th International Workshop on Quantum Physics and Logic (QPL 2011)},
  series    = {EPTCS},
  volume    = {95},
  pages     = {1--14},
  year      = {2012},
}

@article{quirin-tabareau-2016,
  author  = {Kevin Quirin and Nicolas Tabareau},
  title   = {Lawvere-Tierney Sheafification in Homotopy Type Theory},
  journal = {Journal of Formalized Reasoning},
  volume  = {9},
  number  = {2},
  pages   = {131--161},
  year    = {2016},
  doi     = {10.6092/issn.1972-5787/6232},
}

@inproceedings{zeuner-mortberg-2024,
  author    = {Max Zeuner and Anders M\"{o}rtberg},
  title     = {A Univalent Formalization of Constructive Affine Schemes},
  booktitle = {28th International Conference on Types for Proofs and Programs (TYPES 2022)},
  series    = {Leibniz International Proceedings in Informatics (LIPIcs)},
  volume    = {269},
  publisher = {Schloss Dagstuhl---Leibniz-Zentrum f\"{u}r Informatik},
  year      = {2023},
  pages     = {14:1--14:24},
  doi       = {10.4230/LIPIcs.TYPES.2022.14},
}

@article{gogioso-pinzani-2023,
  author  = {Stefano Gogioso and Nicola Pinzani},
  title   = {The Topology of Causality},
  journal = {arXiv preprint arXiv:2303.07148},
  year    = {2023},
}

@article{bareinboim-pearl-2013,
  author  = {Elias Bareinboim and Judea Pearl},
  title   = {A General Algorithm for Deciding Transportability of Experimental Results},
  journal = {Journal of Causal Inference},
  volume  = {1},
  number  = {1},
  pages   = {107--134},
  year    = {2013},
  doi     = {10.1515/jci-2012-0004},
}

@article{bareinboim-pearl-2016,
  author  = {Elias Bareinboim and Judea Pearl},
  title   = {Causal Inference and the Data-Fusion Problem},
  journal = {Proceedings of the National Academy of Sciences},
  volume  = {113},
  number  = {27},
  pages   = {7345--7352},
  year    = {2016},
  doi     = {10.1073/pnas.1510507113},
}

@article{riehl-shulman-2017,
  author  = {Emily Riehl and Michael Shulman},
  title   = {A Type Theory for Synthetic $\infty$-Categories},
  journal = {Higher Structures},
  volume  = {1},
  number  = {1},
  pages   = {147--224},
  year    = {2017},
}

@inproceedings{weaver-licata-2020,
  author    = {Matthew Z. Weaver and Daniel R. Licata},
  title     = {A Constructive Model of Directed Univalence in Bicubical Sets},
  booktitle = {Proceedings of the 35th Annual ACM/IEEE Symposium on Logic in Computer Science (LICS)},
  year      = {2020},
  pages     = {915--928},
  doi       = {10.1145/3373718.3394794},
}

@article{gwb-2024,
  author  = {Daniel Gratzer and Jonathan Weinberger and Ulrik Buchholtz},
  title   = {Directed Univalence in Simplicial Homotopy Type Theory},
  journal = {arXiv preprint arXiv:2407.09146},
  year    = {2024},
}

\end{document}